\documentclass[12pt]{article}
\usepackage[utf8]{inputenc}
\usepackage[a4paper,left=3cm,right=3cm,top=3cm,bottom=4cm]{geometry}

\usepackage{hyperref}
\usepackage{natbib}
\usepackage[ruled]{algorithm2e}
\usepackage{csquotes}
\usepackage{authblk}
\usepackage{amsmath, amssymb, graphicx, multirow,url,hyperref}

\hypersetup{
           breaklinks=true,   % splits links across lines
           colorlinks=false,   % displays links as colored text instead of blocks
        }

\title{30 years of synthetic data}

\author[1,2]{Jörg Drechsler}
\author[2,3]{Anna-Carolina Haensch}	
\date{}
\affil[1]{Institute for Employment Research, Germany}
\affil[2]{The Joint Program in Survey Methodology, University of Maryland, College Park, USA}
\affil[3]{Ludwig-Maximilians-Universität, Munich, Germany}

\begin{document}
\maketitle

\begin{abstract}
  \noindent  The idea to generate synthetic data as a tool for broadening access to sensitive microdata has been proposed for the first time three decades ago. While first applications of the idea emerged around the turn of the century, the approach really gained momentum over the last ten years, stimulated at least in parts by some recent developments in computer science.
  We consider the upcoming 30th jubilee of Rubin's seminal paper on synthetic data \citep{rubin:1993} as an opportunity to look back at the historical developments, but also to offer a review of the
diverse approaches and methodological underpinnings proposed over the years. We will also discuss the various strategies that have been suggested to measure the utility and remaining risk of disclosure of the generated data.
\end{abstract}

%\begin{keyword}
%\kwd{confidentiality}
%\kwd{privacy}
%\kwd{data generation}
%\end{keyword}

%\end{frontmatter}
%%%%%%%%%%%%%%%%%%%%%%%%%%%%%%%%%%%%%%%%%%%%%%
%%%% Main text entry area:

\section{Introduction}\label{sec:intro}
We live in a data-driven world today. Data are collected whenever we interact with our environment, whenever we use our loyalty card in the supermarket, measure our physical activities through wearables, when we check the online weather forecast for our weekend trip, or when we stay in contact with our friends using social media. In the public sector, the ever-growing importance of data is reflected in concepts such as evidence-based policy, and open data movements (see, for example, \citet{EU_open} or \citet{US_open}), and the fact that increasingly more countries explicitly define their own data strategies (see, for example, \citet{EU_data} or \citet{UK_data} for the UK). In industry, the increased reliance on machine learning methods for decision-making results in ever-growing demands for more data to train these models.

However, the increased availability and storage of data also raises concerns regarding confidentiality and privacy. There is an increasing tension between the societal benefits of our digitized world and broad data access on one hand and the potential harms resulting from misuse of data that have not been sufficiently protected on the other hand. Data providers have been concerned about these risks for decades, and various strategies have been developed over the years to avoid disclosing sensitive information when disseminating data to the public \citep{domingo2012,duncan2011}. Still, several prominent examples of confidentiality breaches both in the public and in the private sector \citep{homer2008,ohm2009,sweeney2013,deMontjoye2015} have demonstrated that risks of disclosure often still tend to be underestimated. Increasing computer power and the fact that more and more data are publicly available or are sold by private companies also imply that traditional protection strategies such as swapping, top-coding, or suppression are no longer sufficient to adequately protect the data.

A promising alternative to address the trade-off between broad data access and disclosure protection is the release of synthetic data. With this approach, a model is fitted to the original data and draws from this model are used to replace the original values. Depending on the desired level of protection, only some records (partial synthesis) or the entire dataset (full synthesis) are replaced by synthetic values.

The idea of using synthetic data as a disclosure avoidance strategy is commonly attributed to \cite{rubin:1993} and \cite{little:1993} (although related ideas have been proposed earlier by \cite{liew1985data}). Their approach to synthetic data was motivated by their own work on multiple imputation (MI) for nonresponse \citep{rubin1978,little1987statistical}. Instead of imputing missing values, they suggested adopting the approach to replace sensitive values with \enquote{imputed} values. Similar to the nonresponse context, the release of multiple synthetic datasets would then allow to obtain valid variance estimates that also account for the uncertainty from the synthesis models (assuming the models are correctly specified). However, it took another ten years before the methodology was fully developed, and the first practical applications started to emerge.

Independent of the developments in the statistical community, the computer science community also proposed relying on synthetic data as a way of mitigating the risks of disclosure. The large body of work developed in this field has only rarely aimed at ensuring valid statistical inference, including properly quantifying the uncertainty in the estimates obtained from the synthetic data. The research on synthetic data in computer science was (and still is) predominantly motivated by providing easier data access to train machine learning models. Still, both approaches to synthetic data share the same core goal: ideally, any analysis run on the synthetic data should provide approximately the same answers that would have been obtained if the analysis were run on the original data.

While the body of research has grown steadily over the last thirty years and first deployments of the idea date back to the turn of the century, the concept really gained momentum in the last five to ten years. Many statistical agencies, but also other government agencies such as tax authorities, ministries, or central banks, are exploring synthetic data approaches as a promising tool to broaden public access to their data. Especially within the health sector, the approach gained popularity with applications ranging from generating synthetic patient data \citep{choi_generating_2018} over synthetic electronic health records \citep{yahi2017} to generating synthetic cell images \citep{swicki:2021}. More and more start-ups are offering synthetic data generation as a service and in the industry, synthetic data are used in such diverse contexts as autonomous driving \citep{Osinski2020}, classifying computed tomography images \citep{Fridadar2018} or environmental monitoring \citep{allken_fish_2018}.

Given this growing interest in the field, we consider the 30th jubilee of synthetic data as an opportunity to look back at the historical developments, but also to offer a review of the diverse approaches and methodological underpinnings proposed over the years. We need to emphasize at this point that the diversity of the field and the exponential growth in literature in recent years makes it impossible to offer a detailed review of all methodological tweaks and use cases. We will therefore limit our review to synthetic data methods and applications that specifically aim at offering confidentiality protection. Other contexts in which ideas based on synthetic data have been exploited include, for example, micro simulation \citep{Donoghue2014}, which generates synthetic populations from various data sources, or applications in machine learning, where synthetic data are generated to increase the data pool for model training. Furthermore, we will only discuss and review strategies for synthesis of regular datasets, that is, data structures in which the units are organized in rows and the columns contain the information collected about these units. We will not cover synthesis strategies for text data or images.

The remainder of the paper is organized as follows: In Section \ref{sec:history}, we will provide a brief history of synthetic data. Although the bounds are sometimes blurry, we treat the developments in the statistical community separately from the developments in computer science. The inferential procedures for obtaining valid inferences for the multiple imputation inspired synthesis approaches are covered in Section \ref{sec:mi_inference}. In Section \ref{sec:taxonomy}, we provide a taxonomy for synthetic data. We offer different approaches how to classify various synthesis strategies and also discuss some extensions that have been proposed in the literature. Sections \ref{sec:utility} and \ref{sec:risk} discuss various approaches to measure the utility and remaining risks of disclosure.  The paper concludes with a discussion of verification servers which might help enhance the usefulness of synthetic data in the future.

\section{A brief history of synthetic data}\label{sec:history}
\subsection{The statistical approach}

The idea of releasing synthetic data instead of the real data was first proposed by \citet{rubin:1993}. In a discussion of another article, he suggested that his multiple imputation framework \citep{rubin1978, rubin1987} could be used as an innovative disclosure protection strategy. He proposed treating the units that were not sampled for the survey as missing data and to multiply impute this ``missing'' information. Simple random samples from these imputed populations should then be disseminated to the public. (We note that in practice the two-step procedure of imputing the full population first and then sampling from it is not necessary. It suffices to draw a new sample from the sampling frame and to generate synthetic values for the survey variables of the sampled units.)
If the risk of releasing original records should be avoided completely, the records in the original sample can also be replaced by draws from the imputation model.

Similar to multiple imputation for nonresponse, valid inferences can be obtained from the synthetic datasets by analyzing each dataset separately and combining the estimates from each dataset using simple formulas to come up with the final estimates (see Section \ref{sec:mi_inference} for details).

An obvious advantage of the approach is that no original values are included in the released data (for this reason, this approach has been termed the \textit{fully synthetic data} approach in the literature to distinguish it from the \textit{partially synthetic data} approach described below). Furthermore, synthetic values are generated for units that never participated in the survey. Thus, the level of protection is very high. However, this high level of protection comes at a price. The synthetic data are drawn from a model fitted to the original data, and the quality of the synthetic data directly depends on the quality of that model. Finding a model that reflects all relationships in a complex dataset with hundreds of variables and complicated logical constraints between the variables can be challenging.

A closely related approach that overcomes the limitations of the fully synthetic approach was suggested by \cite{little:1993}. With this approach, only the sensitive records and/or records that could be used for re-identification are replaced with synthetic values. Since some true values remain in the dataset, the approach has been termed the \textit{partially synthetic data} approach. The approach offers some flexibility over the fully synthetic data approach. The agency can decide which part of the data needs to be synthesized. The synthesis can range from synthesizing only some records for a single variable, for example all income values for individuals with an income above a given threshold, to synthesizing all variables, basically mimicking the fully synthetic data approach.

A related idea for generating synthetic data was proposed by \cite{fien:94}. He suggested using a smoothed estimate of the empirical cumulative function of the data and releasing bootstrap samples from this distribution. This approach was further developed in \citet{fienberg:1998}.

Ten years after the initial proposal by Rubin and Little, \citet{raghu:rubin:2001} and \citet{reiterpartsyn} developed the full methodology to enable valid inferences based on fully and partially synthetic data, respectively. Similar to multiple imputation for nonresponse, the multiple synthetic datasets are analyzed separately first and the results from the different analyses are combined using simple combining rules to obtain estimates for the first two moments for the statistic of interest. However, these combining rules differ slightly from the rules in the nonresponse context, and they also differ between full and partial synthesis. In 2012, \citeauthor{reiter2012inferentially} identified another difference between the two synthesis approaches: posterior draws of the model parameters which are necessary for full synthesis (and also in the context of multiple imputation for nonresponse) are not required for partial synthesis. Several years later, \cite{raab2016} developed combining rules for a variant of the fully synthetic approach that can also be used if only one synthetic copy of the original data is available (see Section \ref{sec:mi_inference} for further details).

While early illustrations \citep{reiter:2002a,an:little:2007} and applications \citep{kennickel:1997,abo:stin:06} mostly relied on classical parametric modeling approaches for generating the synthetic data, the suite of modeling strategies has been extended over the years, incorporating ideas from the machine learning literature but also adopting strategies to properly account for the complex sampling designs found in most sample surveys. These will be reviewed in more detail in Section \ref{sec:taxonomy}.

The earliest application of the synthetic data idea dates back to 1997, when the U.S. Federal Reserve Board decided to replace monetary values at high risk of disclosure in the Survey of Consumer Finances with synthetic values \citep{kennickel:1997}. \cite{abowd:2001,abowdwood04} demonstrated the usefulness of the approach for longitudinal, linked datasets using data from the French National Institute of Statistics and Economic Studies (INSEE). The most complex synthetic data product generated so far was first released by the U.S. Census Bureau in 2007: the SIPP synthetic beta \citep{abo:stin:06}, contains synthetic records of the Survey of Income Program Participation (SIPP) linked to administrative records from the Social Security Administration and the Internal Revenue Service. Almost all of the more than 600 variables in this longitudinal dataset are synthesized. Since its first release, the dataset has been updated regularly \citep{benedetto2018}. Another early application was OntheMap, a graphical interface that allows visualizing detailed commuting patterns for the entire United States \citep{abo:ger:08}. This application was the first to offer formal privacy guarantees based on a concept called $(\varepsilon,\delta)$-probabilistic differential privacy, a relaxation of the original definition of differential privacy proposed by \cite{dwork2006calibrating}. Three years later, the U.S. Census Bureau released the Synthetic Longitudinal Business Database \citep{kinney2011,kinney2014}, a partially synthetic copy of the Longitudinal Business Database, which is generated from administrative data at the U.S. Census Bureau and covers all businesses in the United States. The U.S. Census Bureau also uses synthetic data to protect sensitive information in the American Community Survey (ACS) \citep{hawalaacs}. Another large scale synthetic data project was conducted by the Maryland Longitudinal Data System Center (MLDSC), which houses longitudinal education data for the state of Maryland, combining data from various sources. The MLDSC launched the Synthetic Data Project in 2016 sponsored by the Institute of Education Sciences with the goal of facilitating access to this rich source of information \citep{bonnery2019,goldstein2020}. %and in some of the data product Longitudinal-Employer-Household Survey (MAYBE, NEED to CHECK).

Outside the United States, the approach has first been used by Statistics New Zealand to release so-called synthetic unit record files (SURFs) for teaching purposes \citep{forbes2008,keegan2013}. A SURF has also been used more recently as input data for a micro-simulation model that estimates the uptake of and fluency in Te Reo Māori, the language of the Māori people, for various scenarios and policy interventions over the period from 2013 to 2040 \citep{nicholson2021}. The German Institute for Employment Research released a partially synthetic version of one wave of its Establishment Panel in 2011 \citep{dre:jas}. The approach was also adopted to facilitate access to the Scottish Longitudinal Study \citep{nowok2017}. This study links Census data with other sensitive information from health records and death registers. Due to the high sensitivity, access to the data is highly restricted. To prepare their analyses, external researchers can request synthetic datasets that are tailor-made to the research questions the users are trying to answer, that is, the synthetic datasets will always only contain those variables that are needed for the planned research. The R package \textit{synthpop} \citep{nowok2016}, which is now a popular tool for generating synthetic datasets, was also developed as part of this project.

In 2015, a project under the leadership of Statistics Netherlands developed synthetic public use files for the EU Statistics on Income and Living Conditions (EU-SILC) \citep{deWolf2015}. These data, which are available for download at the Eurostat website \citep{EUSILC_data}, are not meant to provide valid statistical inferences. They can be used for training purposes or for developing analysis code while waiting for accreditation to get access to the restricted scientific use files.  More recently, Statistics Canada generated synthetic data, which was used in a Hackathon hosted by Statistics Canada in 2020 \citep{sallier2020}.

Synthetic data are currently at the development stage at several agencies: Examples include the Urban Institute, which is developing synthetic tax data for the Internal Revenue Service \citep{bowen2020,bowen2022synthetic} and the Australian Bureau of Statistics that is currently evaluating synthetic data as a means of broadening access to its microdata \citep{ABS_synthetic}.

Further practical applications have been discussed in the context of protecting data containing detailed geographical information \citep{WangReiter2012,paiva2014,quick2015,quick2018,drechsler2021}, preserving and protecting longitudinal data structures \citep{drechsleraccounting,mitra2020confidentiality}, small area estimation \citep{Sakshaug2010, sakshaug2014_JAS}, synthesizing business data \citep{lee2013regression,drechsler2014PSD,drechsler2014IAOS,alam2020,chien2020synthetic,thompson2022} dealing with nested data structures \citep{hu2018dirichlet} or accounting for complex survey designs \citep{mitra2006,dong2014,dong2014combining,zhou2016,kim2021,hu2022_tabular}. \citet{hurisk} proposed a strategy to reduce the risk of disclosure for partially synthetic data by down-weighting the contribution of high-risk records to the Likelihood function of the synthesizer, while \cite{wei2016releasing} developed a synthesis strategy, which preserves additive constraints. %Additional case studies are presented in \cite{guo2022data,cao2022privacy}.

\subsection{The computer science approach}

% match length and style of section before

The synthetic data approach did not get much attention in the literature on data privacy in computer science before the turn of the century, although \cite{liew1985data} proposed a synthetic data approach for disclosure protection (even though they never called it that way) several years before Rubin's seminal paper. We postulate that this can be attributed (at least in part) to the fact that privacy standards play an important role in the literature on data privacy and the privacy standards used before the advent of differential privacy (DP, \cite{dwork2006calibrating}) only focused on the properties of the data at hand. Popular standards such as $k$-anonymity \citep{Sweeney2002}, $l$-diversity \citep{machanavajjhala2007diversity}, or $t$-closeness \citep{Li2007} all establish certain requirements regarding the properties of the data to consider the data safe from (certain types of) privacy attacks. Synthetic data do not really fit into this notion of privacy. For example, a fully synthetic dataset might not fulfill $k$-anonymity for any $k>1$, but intuitively still offers a high level of privacy protection.

Differential privacy brought a fundamental change in the way computer scientists think about privacy, which paved the way for synthetic data applications in the computer science literature. In a nutshell, differential privacy requires that changing one record in the database has a strictly limited impact on the results of a mechanism run on the data (we will offer a more detailed review of differential privacy in Section \ref{sec:DP_synthesis}). Note the change of focus from the data to the mechanism, which implies that it is no longer the data that needs to be adjusted to achieve privacy, but the mechanism. This concept of privacy aligns much better with the ideas of synthetic data. All that is required is to find a synthesis mechanism that satisfies the requirements of differential privacy. Soon after the concept of DP was established in 2006, the first papers on differentially private synthetic data started to appear.

One of the first approaches was developed by \cite{barak_privacy_2007} who generated synthetic data using a Fourier transformation and linear programming for low-order contingency tables \citep{dwork_2008}. Other early applications include \cite{eno_generating_2008,cano_evaluation_2010,blum_learning_2011, Xiao2011}. Several papers also explicitly adapted the ideas from the statistical community to the DP context \citep{domingo-ferrer_how_2008, abo:ger:08, charest_how_2011, mcclure_differential_2012}. The approach of \cite{abo:ger:08} was later extended in \cite{quick2021generating} and \cite{quick2021improving}. \cite{zhang2014privbayes,zhang_privbayes_2017} proposed an approach that uses Bayesian networks to synthesize high-dimensional datasets, called PrivBayes. In parallel, \cite{li_differentially_2014}  employed Copula functions to take into account the dependency structure of the data (DPCopula).
DP guarantees were also integrated in Generative Adversarial Networks (GANs) later \citep{xie_differentially_2018,yoon_pate-gan_2019}.

The advent of GANs proposed by \cite{goodfellow_generative_2014} resulted in a boost in synthetic data research and applications in the computer science literature. This is probably not surprising as synthetic data are generated as a by-product with any GAN model. We will review GANs in more detail in Section \ref{sec:GAN}, but the basic setup of GANs consists of two neural networks, a generator and a discriminator. The generator produces fake data trying to fool the discriminator, which tries to distinguish the fake data from the real data. Both neural networks are improved in an iterative process. The final data produced by the generator can be seen as a variant of synthetic data. GANs turned out to be extremely successful for image and speech recognition and natural language understanding. Early applications of GANs for data synthesis also focused on generating synthetic images (see, for example, \cite{denton2015deep}). However, the approach was quickly adapted for synthesizing microdata (microdata are often referred to as tabular data in the computer science literature. Thus, many approaches explicitly refer to this in the title of the paper or the labeling of the algorithm to distinguish the approach from other applications that focus on images and other types of data). However, the adoption of GANs for microdata poses additional challenges. Microdata often have categorical variables that are sparse and correlations among variable are often weaker than, for example, relationships between pixels that are located next to each other. The position of variables in a dataset is also only rarely informative for microdata, as the individual records are typically independent. Relationships between variable therefore have to be modeled without the help of any kind of spatial information.

Several of the early applications to microdata only focused on specific types of data, such as time series \citep{esteban2017real,yahi2017} or count and binary data \citep[medGAN]{choi_generating_2018}. tableGAN \citep{park_data_2018} claims to be the first approach capable of handling continuous and categorical variables simultaneously. The approach is built on a GAN originally used for image data by converting records in the original table into a square matrix form. medGAN was extended to categorical variables and further refined in several works \citep{camino_generating_2018,baowaly_synthesizing_2019}.
Later applications relied on Wasserstein-GANs (WGANs)  \citep{koivu_synthetic_2020,zhao_ctab-gan_2021}.
In recent years, more focus has been put on modeling relationships between variables. Conditional tabular GAN (CTGAN) developed by \cite{xu_modeling_2019} addresses challenges from imbalanced categorical and multi-modal continuous data. Causal Tabular GAN \citep[Causal TGAN]{wen_causal-tgan_2021}  allows for modeling the causal relationships between variables in datasets.

Beyond approaches based on GANs other synthesis strategies based on (Variational) Autoencoders \citep{camino_generating_2018,vardhan2020generating,ma_vaem_2020}, Bayesian Networks \citep{zhang_privbayes_2017}, copulas  \citep{patki2016synthetic,kamthe_copula_2021}, or approaches that explicitly preserve certain marginal distributions \citep{mckenna_graphical-model_2019,mckenna_winning_2021} have also been developed in recent years. For a short taxonomy of approaches, see Section \ref{sec:taxonomy}.

The earliest deployment of a differentially private synthesis strategy is OntheMap \citep{abo:ger:08} already mentioned in the previous section. The enforcement of differential privacy for the Decennial Census 2020 can also be seen as a synthetic data approach, as the noisy table counts produced by the TopDown algorithm \citep{abowd2019census} are turned into synthetic microdata from which the actually released tables are generated.

The usefulness of various differentially private synthesis approaches in practical applications was also assessed in the three rounds of the Differential Privacy Synthetic Data Challenge organized by the National Institute of Standards and Technology (NIST) over the years 2018 to 2019 \citep{NIST_challenge}. The winning teams relied on Bayesian networks or approaches to preserve pre-specified marginal distributions. See \cite{bowen2019} for a review of the results from the competition.

Further applications of computer science approaches have been envisioned, proposed and conducted by academic institutions and industry alike. %Industry players that offer synthetic data point towards the greater possibilities for the development, training and testing of AI/ML models through synthetic data and promise safe data sharing both internally and externally.  Main areas of (potential) applications in the commercial and public administration sector due to the sensitive nature of the data are banks and the financial sector, insurances and healthcare and pharmaceutical companies.
GANs for example have been used to create synthetic health records by \cite{esteban2017real,torfi2020privacy} and \citet{Torfi2020}. \cite{chen2019validity} also provides a validation study of a synthetic data generator for patient data with mixed results. Beyond the microdata context that is the focus of this review, GANs have also been used to create realistic images of, for example, skin lesions \citep{pmlr-v116-ghorbani20a}, pathology slides \citep{mahmood2019deep}, and chest X-rays \citep{waheed2020covidgan}.

\section{Obtaining valid inferences for the MI inspired approaches}\label{sec:mi_inference}
As indicated in the introduction, Rubin's initial proposal for data synthesis was motivated by his prior work on multiple imputation for nonresponse. Given the close relationships to those ideas, it seems natural to also adopt the simple combining procedures from the multiple imputation literature (Rubin's combining rules) to obtain valid point and variance estimates from the synthetic data. % that allow making inference regarding the underlying population.
However, the synthetic data approaches differ in two important aspects from the original framework. With full synthesis as proposed by Rubin, synthetic data are only generated for a simple random sample of the population. This extra sampling step needs to be taken into account. With partial synthesis, the synthesis models are estimated using the full data and not only the fully observed subset of the data, as done in the nonresponse context. These deviations imply that the combining procedures also need to be adjusted. The correct rules for fully synthetic data were derived in \cite{raghu:rubin:2001}, those for partially synthetic data are presented in \cite{reiterpartsyn}. Later, \cite{reitermulti} also derived the multivariate analogs that can be used for multi-component testing based on Wald tests or Likelihood ratio tests. We will only review the combining rules for univariate estimates here borrowing heavily from \cite{Drechsler2011}. The interested reader is referred to \cite{reiter:raghu:07}, which offers a full review of all combining rules for synthetic data and also for the nonresponse context.

To understand the procedure of analyzing multiply imputed synthetic datasets,
think of an analyst interested in an unknown scalar parameter $Q$, where $Q$ could be, for example, the mean of a variable, the correlation
coefficient between two variables, or a regression coefficient in a
linear regression. For simplicity, assume that there are no data with items missing in the observed dataset. Inferences for $Q$ derived from the original dataset usually are based on a point estimate $q$, an
estimate for the variance of $q$, $u$, and a normal or Student's $t$
reference distribution. For analysis of the synthetic datasets, let
$q^{(i)}$ and $u^{(i)}$ for $i=1,...,m$ be the point and variance estimates
for each of the $m$ synthetic datasets. The following quantities are
needed for inferences for scalar $Q$:
\begin{eqnarray}\label{q_uni_fully}
\bar{q}_{m} &=& \sum_{i=1}^{m} q^{(i)}/m, \\
b_{m} &=& \sum_{i=1}^{m} (q^{(i)} -
\bar{q}_{m})^{2} / (m-1),\\
\bar{u}_{m} &=& \sum_{i=1}^{m} u^{(i)}/m.
\label{u_uni_fully}
\end{eqnarray}

\subsection{Combining rules for fully synthetic data}

The analyst can use $\bar{q}_{m}$ as an unbiased point estimate for $Q$. Its variance can be estimated using
\begin{eqnarray}
T_{f} =  (1+m^{-1})b_{m} - \bar{u}_{m}.
\end{eqnarray}
When $n$ is large,
inferences for scalar $Q$ can be based on $t$ distributions with
degrees of freedom $\nu_{f} = (m - 1) (1 -\bar{u}_m/((1+m^{-1} )b_{m}))^{2}$.
A disadvantage of this variance estimate is that it can become
negative. For that reason,~\cite{reiter:2002} suggested a slightly
modified variance estimator that is always positive but will tend to overestimate the true variance,
 $T_f^{*}=\max(0,T_f) + \delta(\frac{n_{syn}}{n}\bar{u}_m)$, where $\delta=1$ if $T_f<0$ and $\delta=0$ otherwise.
Here,  $n_{syn}$ is the number of observations in the released
datasets sampled from the synthetic population.

\subsection{Combining rules for partially synthetic data}

Similar to fully synthetic data, the analyst can use $\bar{q}_{m}$ to estimate $Q$. The variance of $\bar{q}_{m}$ for partially synthetic data can be estimated using
\begin{eqnarray}
T_p=b_m/m+\bar{u}_m.
\end{eqnarray}

When $n$ is large, inferences for scalar $Q$ can be based on
$t$ distributions with degrees of freedom
$\nu_p=(m-1)(1+\bar{u}_m/(b_m/m))^2$.
Note that the variance estimate $T_p$ can never be
negative, so no adjustments are necessary for partially synthetic datasets.

\subsection{An alternative variance estimate for fully synthetic data}
When generating fully synthetic data, most researchers do not follow the protocol as initially envisioned by \cite{rubin:1993}. Rubin assumed that in addition to the survey variables $Y$ some additional variables $X$ would be available for the full population. In the survey context, these variables represent design variables available from the sampling frame. Under this assumption, fully synthetic data for $Y$ would be generated by fitting a model for $f(Y|X)$ using the survey data and using this model to generate synthetic values for a new sample of design variables $X^{new}$ by drawing from $f(Y|X^{new})$. Only the synthetic $Y$ values would then be released to the public.

In practice, researchers typically only use the information in $Y$ to generate synthetic data. In this setting, fully synthetic data can be seen as an extreme variant of partial synthesis for which the set of unsynthesized records is empty. This also implies that the combining rules for partial synthesis are still valid as first noted by \cite{drechsler2011fully}. Extending these ideas, \cite{raab2016} proposed an alternative variance estimator that can be used in this situation:
\[
T_s=\left(\frac{n_{syn}}{n_{org}}+\frac{1}{m}\right)\bar{u}_m,
\]
where $n_{syn}$ is the number of synthetic records and $n_{org}$ is the number of records in the original dataset. Note that this variance estimator does not rely on the between imputation variance $b_m$. This offers three important advantages compared to $T_f$, the variance estimator for fully synthetic data discussed above: (i) the estimator $T_s$ can never be negative, (ii) it has less variability than $T_f$ ($b_m$ is only an estimate for the true variability between the datasets and the fact that it is based on a limited number of $m$ synthetic datasets implies high uncertainty in this estimate, which is also the reason why $T_f$ can sometimes be negative), (iii) valid variance estimates can be obtained from a single synthetic dataset. The last point is especially important because previous research has shown that the risk of disclosure increases with the number of synthetic datasets \citep{dre:rei:jos,reiter2010}. Of course, the price to pay is an increased level of uncertainty, if only one synthetic dataset is released. Note that assuming $n_{syn}=n_{org}$, the variance can be reduced by 25\% when releasing two datasets instead of one dataset. These accuracy gains are diminishing quickly with increasing $m$ and relative reduction in variance is bounded by 0.5 for $m\rightarrow\infty$. See \cite{drechsler2018} for further discussion of the advantages and disadvantages of the different synthesis strategies and which variance estimator is appropriate in which scenario.

\section{A taxonomy of synthetic data approaches}\label{sec:taxonomy}
Given the broad range of synthetic data approaches and use cases, finding a one-dimensional taxonomy that fully covers all variants of synthetic data is difficult. Beyond the obvious distinction between approaches inspired by the ideas of multiple imputation and approaches that originated in computer science, we suggest three alternative classification schemes: sequential vs. joint modeling approaches, parametric vs. machine learning inspired approaches and approaches that offer formal privacy guarantees vs. those that do not. Obviously, other classifications such as partial vs. full synthesis would be possible. However, we feel that classifying the approaches along these lines is obvious and does not require a separate discussion. Instead, we list a final class of synthesis approaches that are extensions of the MI based approaches. These approaches are treated separately, as they typically require different procedures to obtain valid inferences compared to those discussed in Section \ref{sec:mi_inference}.

\subsection{Sequential vs. Joint Modeling}
Most of the early applications of synthetic data relied on a sequential modeling approach, in which each variable is synthesized sequentially using models that condition on any variables that have been synthesized previously or variables that remain unchanged in the final data. The underlying idea is that any joint distribution can be rewritten as a product of conditional distributions. %To illustrate, let $X$ denote those variables that will not be synthesized ($X$ will be empty for full synthesis) and let $Y=\{Y_1,Y_2,Y_3\}$ denote three variable to be synthesized. We can write the joint distribution of the data as $f(Y,X)=f(X)f(Y_1|X)f(Y_2|Y_1,X)f(Y_3|Y_1,Y_2,X)$. The sequential regression approach would generate synthetic data in three steps: (i) fit a model for $f(Y_1|X)$ and generate synthetic values $Y_1^{(syn)}$ by randomly drawing new values from that model. (ii) use the original data to fit a model for $f(Y_2|X,Y_1)$ and generate synthetic values $Y_2^{(syn)}$ by drawing from $f(Y_2|X,Y_1^{(syn)})$. (iii) use the original data to fit a model for $f(Y_3|X,Y_1,Y_2)$ and generate synthetic values $Y_3^{(syn)}$ by drawing from $f(Y_3|X,Y_1^{(syn)},Y_2^{(syn)})$. Note that while the model parameters are always estimated using the original data, synthetic values are generated conditioning on the previously synthesized variables.

The sequential regression approach offers great flexibility, as different models can be used for each variable. These might include parametric models such as linear regression, logit models \citep{reiter:2002a}, or models based on quantile regressions \citep{pistner2018synthetic}, but also any machine learning tool that enables random draws from a conditional distribution, such as CART or random forests.

In contrast to the sequential modeling approach, joint modeling aims at directly specifying the joint distribution of the data. While early approaches such as the IPSO method \citep{burridge2003} relied on a multivariate normality assumption that is seldom justified with real data, more flexible models have been proposed recently. For categorical data, \cite{HuReiterWang2014} demonstrated that an approach based on a Dirichlet Process Mixture of Products of Multinomials (DPMPM) can offer high utility in real data applications. The approach was later extended to also allow for structural zeros, that is, impossible variable combinations such as married toddlers \citep{manrique2018}. Synthesis approaches based on DPMPMs are implemented in the R package NPBayesImputeCat \citep{hu2021multiple}. A related approach based on Quasi-Mutinomial distributions was proposed by \citet{hu2018quasi}, while \citet{jackson2022PSD} and \citet{jackson2022using} proposed saturated count models for easy synthesis of large databases with a-prori utility guarantees. \cite{kim2018} showed good performance of Dirichlet Process Normal Mixture Models for synthesizing continuous business data. This approach was later extended to also account for informative sampling designs that are common with business surveys \citep{kim2021}. Furthermore, many of the synthesis strategies used in computer science such as Generative Adversarial Networks\citep{goodfellow_generative_2014} or Bayesian Networks \citep{zhang_privbayes_2017} can be subsumed under this category. We will review the literature based on these approaches in Section \ref{sec:CS_approaches}.

\subsection{Parametric vs. ML based}
The methodology for obtaining valid inferences based on synthetic data reviewed in Section \ref{sec:mi_inference} above relies on the assumption that the synthesis models are correctly specified, that is, they match the true data generating process. An additional requirement is that the synthesis model is congenial to the analysis model to be run on the synthetic data. In broad terms, congeniality \citep{meng:1994} means that the synthesis model is based on the same (modeling) assumptions as the analysis model.

%Since many of the early synthetic data applications assumed that the data would be used by social scientists, economists or medical researchers, who typically rely on parametric regression models to understand the relationships between the variables, it seemed natural to also rely on parametric models when generating the synthetic data. Thus, using the sequential modeling approach described above, continuous variables were typically synthesized using linear regression models (potentially after applying necessary transformations to satisfy the normality assumptions), binary variables were synthesized using logit or probit regressions, multinomial regression models were used for variables with more than two categories etc. \cite{reiter:2002a} reviews these synthesis techniques.

To be fair, as it is impossible to anticipate all analyses that will be run on the synthetic data, achieving congeniality is typically a hopeless goal in practice. Still, it has been shown in the nonresponse context  \citep{meng:1994} that approximately valid inferences can be obtained if the synthesis model encompasses the analysis model, that is, it contains more variables than the analysis model. Intuitively, this makes sense: adding a predictor variable during synthesis that in reality is conditionally independent of the variable to be synthesized given the other predictors in the model will not do much harm. It might unnecessarily increase the variance from synthesis, but it will not introduce any bias. However, omitting important variables will introduce bias, as the relationship between the omitted variable and the synthetic variable will be attenuated in the synthetic data.

Based on this reasoning, it is generally recommended to use a rich set of predictors in the synthesis model, ideally conditioning on all other variables in the dataset and also including interaction and squared terms if possible (see \cite{little1997} for a similar argument in the nonresponse context). However, this strategy is typically not feasible when using parametric models, as many datasets contain dozens of variables. Especially with categorical variables, multicollinearity issues and the problem of perfect prediction often imply that fitting parametric models containing many variables is no longer possible and uncongeniality becomes a major concern% unless the assumption that all variables that are not included in the synthesis model are independent of the variable to be synthesized conditional on the variables included in the model is justified
.

To overcome this problem, researchers started exploring alternative synthesis strategies, borrowing ideas from the machine learning literature. In 2005, \cite{reitercart} suggested using Classification and Regression Trees (CART). \cite{rei:RF} later extended these ideas to random forests, and \cite{drechsler2010} developed strategies to adapt Support Vector Machines for data synthesis. Synthesis strategies based on genetic algorithms were explored in \cite{chen2016genetic} and \cite{chen2018application}. All these approaches have the advantage that they \enquote{let the data speak for themselves}, that is, they might automatically identify higher order relationships that might easily be missed when specifying parametric models. Furthermore, they are not affected by multicollinearity or perfect prediction problems and can still be directly applied if the number of variables exceeds the number of observations. In an evaluation study, \cite{dre:rei2011} compared the different approaches and found that CART models offered the best results in terms of preserving the information from the original data. %One possible explanation could be that with synthetic data, the risk of overfitting, arguably the major disadvantage of CART models in other contexts, is less problematic (at least from a utility perspective), as the goal is only to replicate the original data as closely as possible.

In the computer science approach to synthetic data, the problem of uncongeniality was never explicitly considered. Since from the beginning the expected use case was the training of machine learning models, the focus of the research was on machine learning models right from the start.
Before we review the different approaches from the computer science literature, we briefly discuss some extensions of the MI based synthesis procedures.

\subsection{Extensions of the MI inspired approaches}

The approaches reviewed in this section offer various extensions to the classical synthesis problem. They differ from the other approaches in that they require different
inferential procedures than those discussed in Section \ref{sec:mi_inference}. % if the goal is to obtain valid point and variance estimates relative to a population parameter of interest.
We will not review all these procedures here for brevity. Instead, we refer to the various papers for further detail.

The first extension of the classical MI based synthesis approach was proposed by \cite{reitermi}. The paper offers a strategy to deal with missing data and data confidentiality simultaneously. The author proposes a two-step procedure, in which missing values are imputed $m$ times at the first stage and $r$ partially synthetic datasets are generated at the second stage within each first stage nest, that is, the final data comprises $m\cdot r$ datasets. The appropriate procedures for mutli-component hypothesis testing under this scenario were derived in \cite{kinney2010tests}.

In a similar spirit, \cite{reiter2010} proposed a two-stage synthesis, for which variables that have a higher risk of disclosure are synthesized at the first stage and variables that require a larger number of synthetic datasets to reduce the model uncertainty are synthesized on the second stage. This approach was motivated by previous findings \citep{dre:rei:jos} that increasing the number of synthetic datasets can lead to increased risks of disclosure. The authors show that their approach offers better disclosure protection and similar utility compared to standard one-stage synthesis with the same total number of synthetic datasets.

A final type of extension proposes to use a (sub)sampling step before the synthesis. This approach is especially attractive for Census data, for which it is common practice that only random samples of the full data are released to the public. What makes this approach special in the synthesis context is that the synthesis models can be estimated using the full data even if only a (sub)sample is synthesized later. \cite{DrechReit10} present the methodology if the original data covers the full population. Using a real dataset, they illustrate that releasing synthetic samples can actually offer higher utility than releasing samples of the original data. This surprising result is due to the fact that the synthesis models are based on the information from the full population. \cite{drechsler2012} extend the methodology to the context where the original data is itself already a sample.

\subsection{Computer Science approaches}\label{sec:CS_approaches}

In computer science, machine learning and deep learning methods such as Generative Adversarial Networks  \citep[GANs]{goodfellow_generative_2014} and Variational Autoencoders  \citep[VAEs]{kingma_auto-encoding_2014} have been popular generative modeling frameworks in recent years.  Thus, it is perhaps not surprising that a large body of work on synthetic data in computer science is based on one of these concepts. %However, some other synthesis strategies have also been explored.
In this section, we offer a brief overview of the most popular variants of these two approaches. Due to the large body of work in the field, we discuss only the most influential contributions, excluding works that are targeted towards very narrow areas of application.% Congeniality has not been discussed or been an explicit goal of data synthesis in computer science. The evaluation of the utility of generated datasets is therefore done with methods that differ from statistics (see also section REF) and contribute to the exploration of methods that have not been considered in statistics (and vice versa).

%In the following, we provide an overview of developments in computer science. At the beginning of this section, we want to place two notes. First, note that due to the vast number of approaches, we discuss only the most influential works, and we exclude works that are targeted towards very narrow areas of application.
%Second, the literature in computer science often uses the term tabular data, and its significance differs from the one in statistics. In statistics, tables are a form of summarizing findings by groups, but the term \textit{tabular data} in computer science does not refer to this type of summaries. In computer science, tabular data are data that are structured in a tabular form, in general with variables in columns and individual data in rows, i.e. tabular data in computer science corresponds to microdata or individual person data in other fields.

%We will provide an overview of approaches and a small foray into differential privacy-preserving data synthesis at the end. We start with generative adversarial networks (GANs) and then continue with Variational Autoencoders (VAE) before concluding with approaches that offer differential privacy guarantees.  Compared with the abundance of literature on GANs and other deep learning approaches for text, audio and visual data generation, literature on the use of machine learning and especially deep generative learning approaches for tabular data is relatively sparse but rapidly growing.

\subsubsection{Generative adversarial networks (GANs)}\label{sec:GAN}
 As indicated in the introduction, we will only focus on GANs for microdata synthesis in this review. Compared with the abundance of literature on GANs and other deep learning approaches for text, audio and visual data generation, literature on the use of deep generative learning approaches for synthesis of microdata is relatively sparse but rapidly growing
%In recent years, a multitude of methods based on  general adversarial networks (GANs) have been developed
\citep{park_data_2018,choi_generating_2018,camino_generating_2018,koivu_synthetic_2020,xu_modeling_2019}.

GANs consist of two neural networks that compete with each other: the so-called generator (network) is trained to generate synthetic data and outputs synthetic samples given a random noise input. The discriminator (network) is trained to discriminate between real and synthetic data. The discriminator tries to minimize the misclassification error %penalizes itself for misclassifying a real dataset as synthetic, or a synthetic dataset created by the generator as real. In comparison,
 while the generator loss is calculated from the discriminator’s classification – it gets penalized if it does not fool the discriminator. The standard combined loss function was described by \cite{goodfellow_generative_2014} and is also called minimax loss, since the generator tries to minimize it while the discriminator tries to maximize it.  The training of the GAN is an iterative process in which each of the neural networks updates its parameters based on the feedback received from the other network, that is, GANs make use of adversarial feedback loops to learn how to generate synthetic data that is indistinguishable from real data.

%While GANs have been extremely popular for the creating of synthetic images, there has also been a corpus of work %\cite{park_data_2018,choi_generating_2018,camino_generating_2018,koivu_synthetic_2020,xu_modeling_2019}
%that explore using GANs for the generation of microdata.
%We will give an overview of approaches and group them by the exact methods that they employ.

In recent years, Wasserstein GANs (WGANs) \citep{arjovsky_wasserstein_2017} have become increasingly popular.
WGANs use the Wasserstein distance for the cost function instead of the  Kullback-Leibler (KL) and Jensen–Shannon (JS) Divergence to avoid the problem of vanishing gradients %by having a smoother gradient
\citep{arjovsky_wasserstein_2017}.  The Wasserstein distance is also called Earth Mover's distance and is widely used to solve optimal transport problems, that is, problems where the goal is to move things from a given configuration to a desired configuration with the smallest cost possible. Early examples for the use of WGANs for data synthesis can be found in \cite{camino_generating_2018}.

However, for WGANs, the discriminator usually has to obey a Lipschitz constraint. To enforce this constraint, the weights of the discriminator must be within a certain range controlled by a hyperparameter. \cite{arjovsky_wasserstein_2017} propose to clip the weights if necessary, but noted that this approach is not optimal. To overcome this problem, WGAN-gradient penalty (WGAN-GP) \cite{gulrajani_improved_2017} uses a gradient penalty to fulfill the Lipschitz constraint.  \cite{baowaly_synthesizing_2019},\cite[actGAN]{koivu_synthetic_2020}, \cite{xu_modeling_2019} and \cite[CTAB-GAN]{zhao_ctab-gan_2021} all use different adaptations of WGAN or WGAN-GP for data synthesis. \cite{xu_modeling_2019} use normal mixture distributions to improve the fit for continuous variables. They also use a conditional generator, aiming for proper conditional distributions for each variable.

There also exist alternatives to WGANs for data synthesis, for example, GANs
based on the Cramér Distance \citep{mottini_airline_2018}.

Causal-TGAN is an approach that stands out from other GAN approaches, as it explicitly takes the potentially complex causal relationships between the variables into account. It is composed of two steps, first obtaining the causal graph that represents the causal relations of the original dataset and then using the causal graph when training the GAN \citep{wen_causal-tgan_2021}.

%With a similar focus on representing relationships between variables, \cite{xu_modeling_2019}  propose to normalize multimodal or skewed data by expressing  absolute values as the combination of a previously-identified mode and the deviation of the value from that mode. They also use a conditional generator, aiming for proper conditional distributions for each variable.

\subsubsection{Variational Autoencoders (VAEs)}

Another approach based on deep neural networks that has been adapted for data synthesis lately are variational autoencoders \citep[VAE]{kingma_auto-encoding_2014}.
In comparison to GANs, a VAE has three instead of two networks, which learn complimentary tasks: an encoder network, a decoder network and a discriminator. The encoder network maps the data onto a latent representation,  while the decoder network tries to reconstruct. As with GANs, the discriminator network decides for each given sample whether it is real data or data generated by the decoder network. A VAE is trained to minimize the reconstruction error between the reconstructed data and the initial data.
 Data synthesis approaches that use VAE are discussed in \citet[VEEGAN]{srivastava_veegan_2017} \cite{camino_generating_2018,xu_modeling_2019,vardhan2020generating}, and \cite{ma_vaem_2020}.

\subsection{Differential private data synthesis}\label{sec:DP_synthesis}
In recent years, differential privacy \citep[DP]{dwork2006calibrating} has been widely adopted as a definition of privacy offering formal, that is, mathematically quantifiable privacy guarantees. DP requires that the impact that any single record can have on the probability of obtaining a specific result is strictly bounded. Specifically, pure $\varepsilon$-DP requires that the log-difference in the probability of obtaining a specific output computed on two neighboring datasets, that is, datasets that differ only in one record, is bounded between $\epsilon$ and $-\epsilon$. In layman's term, an algorithm is differentially private if someone seeing the output statistic cannot tell if the information on a specific individual was used in the computation or not. See \cite{dwork2014} or \cite{vadhan2017complexity} for an in-depth discussion of differential privacy and some relaxations of the concept that have been proposed in the literature. The body of work on DP has grown exponentially in recent years and several tech companies, such as Apple \citep{adp2017}, Google \citep{erlingsson2014}, and Microsoft \citep{ding2017} as well as the U.S. Census Bureau \citep{Foote2019, abowd2022} recently adopted the approach for some of their data. %In short, if a statistic is released after using a differentially private mechanism, the log-difference on the probability to obtain a specific value of the statistic calculated in two neighboring data sets, i.e. differing by one recorded individual or unit, is bounded between $\epsilon$ and $-\epsilon$. In layman's term, an algorithm is differentially private if someone seeing the output statistic cannot tell if information on a specific individual was used in the computation or not.
%The concept of DP has spurred a huge amount of work in differentially private mechanisms for general settings, for specific statistical analyses as well as software and web-interfaces.

The concept of DP has also stimulated research on generating differentially private synthetic data.
Differentially private synthetic data have an advantage over other approaches for private data analysis: DP is \textit{immune to post-processing}, that is, any function of a differentially private output is guaranteed to also be differentially private with the same privacy guarantees as the original output. This implies that researchers working with the differentially private synthetic data are more free to interact with the data and use any tools and workflows to process the data without the risk of accidentally or purposefully revealing any sensitive information.

Various approaches have been proposed in the literature for generating differentially private synthetic data (see %also called Differentially Private Data Synthesis (DIPS)
\cite{bowen_comparative_2020} for a review of early approaches).
Using marginal distributions for the synthesis has been one of the most popular approaches. Noise is added to either one-, two- or three-way marginal distributions \citep{mckenna_graphical-model_2019,mckenna_winning_2021,liu2021leveraging}. %A central challenge to these approaches is to preserve correlations between ariables while still guaranteeing privacy.
%In recent years, many organizations and institutions like the 2020 U.S. Census (Abowd, 2018) or companies like  Microsoft (Ding et al., 2017) and Apple (Differential Privacy Team, Apple, 2017) have begun to use differential privacy \cite{dwork_differential_2006}, as a mathematically rigorous measure of privacy when providing statistics.
%Differentially private synthetic data sets have therefore been proposed as an alternative.  Because synthetic data already meets the requirements of differential privacy, researchers are more free to interact with the data.  There has been a wide range of  approaches applied to the synthesis of microdata. The winning group from the National Institute of Standards and Technology (NIST) challenge \cite{mckenna_graphical-model_2019,mckenna_winning_2021} used noisy three-way marginals for data synthesis.
Another  popular approach for differentially private data synthesis are Bayesian networks  \citep{bao_synthetic_2021}, most prominently PrivBayes by \cite{zhang_privbayes_2017}. It can be difficult to represent all important correlations in PrivBayes.  Therefore, \cite{cai_data_2021} propose a Markov random field (MRF) that models the correlations among the variables in the original datasets, and then uses the MRF for data synthesis (PrivMRF).
Game-based approaches such as those by  \citet[MWEM]{hardt_simple_2012} and \citet[Dual-Query]{gaboardi14}  require a set of specified queries, optimizing the synthesis to ensure high validity for these queries. Yet another popular approach developed by \citet[DPCopula]{li_differentially_2014} is based on Copula functions.  %However, aiming for greater computational efficiency and towards applications with higher-dimensional data problems, \cite{liu2021leveraging}  incorporate public data into MWEM.
%An overview of a number of other (early) DP-data synthesis methods is given by \cite{bowen_comparative_2020}.  %

Finally, work on integrating differential privacy into generative adversarial networks (GANs) has been growing fast in the last few years \citep{beaulieu-jones_privacy-preserving_2019,xie_differentially_2018, yoon_pate-gan_2019,torkzadehmahani_dp-cgan_2020,neunhoeffer_private_2021}. Since the generator commonly never accesses the real data directly, only the discriminator needs to be modified to ensure DP:
\cite{beaulieu-jones_privacy-preserving_2019} and \cite{xie_differentially_2018} built on \cite{abadi_deep_2016} for the private optimization, adding Gaussian noise to the gradient of the Wasserstein distance in the WGAN algorithm. The gradients are also clipped if necessary. %These steps are done in order to obfuscate the influence that the private input data can have on the final model for creating differentialy private data.
\cite{frigerio_differentially_2019} also proposes a private extension of WGAN.  Conditional GANs \citep[CGAN]{Mirza2014} are adapted by \cite{torkzadehmahani_dp-cgan_2020}.  % but using a privacy accountant to compute the overall privacy costs by calculating the cost of a single iteration of the algorithm and cumulating it with the other iterations.
%Again, similar to the previous two approaches, \cite{torkzadehmahani_dp-cgan_2020} adapts an existing GAN approach by adding Gaussian noise to the accumulation of gradients of discriminator on real and fake data, but in their case adapting CGAN not WGAN.
\cite{yoon_pate-gan_2019} use the Private Aggregation of Teacher Ensembles (PATE) framework proposed by \cite{papernot_scalable_2018}, which provides a differentially private method for classification tasks. The framework is used for the discriminator's task to differentiate real and fake data.

To provide greater robustness against low utility of generated DP data sets,  \cite{neunhoeffer_private_2021} proposed a method combining weighted samples produced by a sequence of generators. Their approach can be applied to different private or non-private GANs for data synthesis.

%\paragraph{Specialized/unsolved problems evtl.& Fazit}

%\begin{itemize}
%        \item Fehlendenaten ITS-GAN
%        \item hierarchischen Datensätze canale et al 2022/Generating Synthetic Data for Multiple Tables
%        \item Preserve constraints in data
%        \item Need for Robust Evaluation Frameworks
%\end{itemize}

\section{Utility Evalution}\label{sec:utility}
There is a large body of literature on measuring the validity of data that has undergone some form of perturbation to protect confidentiality. Most of these methods can also be used to measure the validity of synthetic data. We will focus on the measures that are most relevant for synthetic data. Additional measures are discussed, for example, in \cite{domingo2012} or \cite{arnold2020really}.

Utility metrics can be broadly divided into three categories: The first category, commonly referred to as \textit{global utility metrics}, tries to assess the utility by directly comparing the original data with the protected data. These measures offer the advantage that no assumptions regarding the types of analyses the synthetic data will be used for need to be made. On the downside, given that utility is measured on a very aggregated level, good results for these measures do not necessarily guarantee high utility for a specific type of analysis the user might be interested in. \textit{Outcome-specific utility metrics} sit on the other end of the spectrum. They measure the utility for a specific analysis, for example, the results of a linear regression model. A third class of measures that we label \textit{fit-for-purpose measures} usually form the starting point of any utility assessment. Examples of these measures would be graphical comparisons of the marginal and bivariate distributions of all variables or consistency checks to avoid implausible values such as negative age values in the synthetic data.

\subsection{Global utility metrics}
As discussed above, these measures try to evaluate the utility by directly comparing the synthetic data to the original data. One common approach in this context is to use some distance measure, such as Kulback-Leibler divergence \citep{utility} or Hellinger distance \citep{Gomatam2003}. A downside of these general distance measures is that they can be difficult to compute for large datasets. An alternative strategy tries to assess how easy it is to discriminate between the original data and the synthetic data, borrowing ideas from the literature on propensity score matching \citep{rosenbaum1983}. Propensity scores are estimated by stacking the $n_{org}$ original records and the $n_{syn}$ synthetic records and adding an indicator, which is one if the record is from the synthetic data and zero otherwise. In the next step, a model is fitted using the information contained in the data to estimate the propensity scores, that is, to estimate the probability for each record to belong to the synthetic data. If the synthetic data would be an exact copy of the original data, the data would not offer any information to discriminate between the data sources and the distribution of the estimated propensity scores would be the same for both datasets. Thus, one way to measure the utility of the synthetic data is to evaluate the difference in the distribution of the propensity score between the original data and the synthetic data. Various metrics can be used for this purpose. \cite{bowen2021} suggest estimating the Kolmogorov-Smirnov distance between the two distributions (they call this measure SPECKS for Synthetic data generation; Propensity score matching; Empirical Comparison via the Kolmogorov-Smirnov distance). Alternatively, the Mann–Whitney U test (Wilcoxon rank-sum test) can also be used.

A measure that gained popularity in recent years is the propensity score mean squared error ($pMSE$) as an evaluation metric \citep{woo,snoke2018}. Let $p_i$, $i=1,\ldots,N$ with $N=n_{org}+n_{syn}$ denote the predicted value obtained from the model for record $i$ in the stacked dataset. The $pMSE$ is calculated as $1/N\sum_N(p_i-c)^2$, with $c=n_{syn}/N$. % The procedure consists of the following steps:
%\begin{enumerate}
%    \item Stack
%    \item Use a model to predict the data source (original/synthetic) using the information contained in the data. Let $p_i$, $i=1,\ldots,N$ with $N=n_{org}+n_{syn}$ denote the predicted value obtained from the model.
%    \item Calculate the $pMSE$ $1/N\sum_N(p_i-c)^2$, with $c=n_{syn}/N$.
%\end{enumerate}
The smaller the $pMSE$ the higher the analytical validity of the synthetic data (note that $p_i\rightarrow c$  if the model cannot discriminate between the original data and the synthetic data). A downside of the $pMSE$ noted by \cite{woo} is that it increases with the number of predictors included in the propensity model. To overcome this problem, \cite{snoke2018} derived the expected value and the standard deviation of the $pMSE$ under the null hypothesis that the synthesis model is correctly specified and proposed two additional utility measures. The first measure is the $pMSE$ ratio which is computed as the empirical $pMSE$ divided by its expected value under the null. The second measure is the standardized $pMSE$, which is the empirical $pMSE$ minus its expectation under the null divided by its standard deviation under the null. In a recent paper, \cite{drechsler2022challenges} critically discussed the $pMSE$ illustrating that the estimated scores are highly dependent on the specification of the propensity model and that even baltant differences in the utility between different synthesizers are sometimes not picked up by the $pMSE$.

\subsection{Outcome-specific utility measures}\label{sec:global_utility}
These measures explicitly focus on measuring the usefulness of the synthetic data for a specific analysis task. For example, a straightforward visualization of the analytical validity is to plot estimates of interest (means, regression coefficients, etc.) obtained from the original data against the same estimates obtained from the synthetic data. If the utility is high, the coefficients should cluster around the 45 degree line.

A downside of this evaluation is that it does not account for the inherent uncertainty of the estimates. Larger deviations between the estimates might be acceptable, if the sampling error is large, for example, if the estimate of interest is based on a small subset of the data. The same deviation might be problematic for a statistic based on the entire sample. A popular measure that also takes the uncertainty of the estimates into account is the \textit{confidence interval overlap} measure proposed by \cite{utility}. It measures the relative average overlap between the confidence interval obtained from the original data and the confidence interval obtained from the synthetic data. An overlap measure close to one indicates that approximately the same inferential conclusions will be drawn irrespective of whether the synthetic data or the original data were used for the analysis.

Given the increased relevance of machine learning approaches, another utility metric gained popularity in recent years, especially in the computer science literature: \emph{machine learning efficacy}. Utility measures of this type, which are also referred to as measures of \emph{model comparability}, assess whether machine learning models trained on the synthetic data give similar results compared to when they were trained on the original data. For these evaluations, the models of interest are typically trained on both the synthetic data and the original data and then the performance of the models is compared based on the same set of test records, which is obtained from the original data. The utility of the synthetic data is considered high, if classical evaluation criteria such as accuracy, F1 score, etc., are similar irrespective of whether the models were trained using the original data or the synthetic data. Sometimes, utility is also evaluated by assessing whether using the synthetic data for model training would lead to the same ranking of various machine learning models. For example, if the original data would suggest that a classifier based on a multilayer perceptron performs better than a random forest and the random forest is better than logistic regression, the same ranking should be found if the synthetic data were used for model training.

\subsection{Fit-for-purpose measures}
These measures represent the first step when evaluating the usefulness of the generated data. We treat them separately from the other two measures, as they do not necessarily focus on measuring the validity of specific analyses that might be important for the users of the data. They also do not try to directly assess the similarity of the original and the synthetic data in one global statistic. Their main aim is to get a first impression of the quality of the synthetic data, and, unlike the global measures, they can help to identify aspects of the synthesis process that might still need to be improved.
These measures can be divided into three groups: graphical evaluations, plausibility checks, and computing various goodness-of-fit measures.

Graphical evaluations typically include strategies such as side-by-side plots of the marginal distributions of the synthetic and the original data or contour plots for comparing bi-variate distributions. They also include visual comparisons of conditional distributions such as the income distribution for males and females or for different age groups.

For the plausibility checks, it is important to involve subject-matter experts that regularly work with the data. This is crucial as not all inconsistencies are immediately obvious. For example, while it might be straightforward to identify problems such as married two-year olds, it is much more difficult to judge which year-to-year change in turnover would be considered plausible for an establishment in a given industry in a given year.

Finally, any goodness-of-fit measure can be used to assess the similarity for specific aspects of the original and synthetic data. For example, the Kolmogorov-Smirnov test statistic can be used for each continuous variable in the dataset. Cross-tabulations of several variables (discretizing continuous variables if necessary) can be evaluated using the $\chi^2$ statistic or the likelihood ratio statistic. \cite{voas2001evaluating} discuss the advantages and disadvantages of various metrics. However, it must be noted that the statistics should not be used to test for statistically significant differences between the original and the synthetic data. Given that the synthetic data are generated based on information from the original data, the two samples cannot be treated as independent---an assumption underlying most goodness-of-fit tests. Thus, any p-values computed using the standard test procedure would be misleading. Nevertheless, the value of the test statistic can still be used to compare the performance of different synthesis strategies. Furthermore, the test statistic can also be used as a metric to identify potential problems with the quality of the synthetic data. For example, if the test statistic is high for many of the cross-tabulations involving age, this serves as an indicator that the synthesis of the age variable needs to be improved.

The $pMSE$ measure discussed in Section \ref{sec:global_utility} can also be used as a fit-for-purpose measure by only including the variables of interest when estimating the propensity score. An illustration of how this strategy can be used to visualize the utility for bi-variate distributions is presented in \cite{raab2021}. These graphical visualization tools are also implemented in the R package synthpop \citep{raab2016}.

In \cite{raab2021}, the authors empirically evaluate various goodness-of-fit measures and find a large correlation ($>0.9$) between most of them. Noticeably, the adjusted $\chi^2$ test proposed by \cite{voas2001evaluating}, the Freeman-Tukey statistic, the Jensen-Shannon divergence (JSD), and the $pMSE$ had an empirical correlation above 0.99, so did the Kolmogorov-Smirnoff test statistic, the Mann-Whitney test statistic, and two additional measures that we don't review here for brevity. In practice, this seems to imply that it is sufficient to only use one or two goodness-of-fit criteria when assessing the utility of the generated data.

\section{Risk Assessment}\label{sec:risk}
From a risk perspective, there is a fundamental difference between disseminating partially or fully synthetic data. With partial synthesis, there still exists a one-to-one mapping between the original data and the synthetic data. With fully synthetic data, this is no longer the case. In fact, with this approach, the synthetic data does not have to be of the same size as the original data. This implies that measuring the risk of re-identification, as commonly done for other disclosure protection strategies \citep{reiter2005,skinner2008,shlomo2014}, is not meaningful for fully synthetic data. However, this does not mean that fully synthetic data can be assumed to have no risk of spilling sensitive information. For example, \cite{manrique2018} illustrate using real data that if a fully conditional specification approach (which is commonly applied when using multiple imputation in the nonresponse context) is used for CART-based synthesis, there is a risk that the synthesizer simply replicates most of the original records.  The problem arises as the approach always conditions on all other variables in the dataset. With complex datasets containing many (categorical) variables, this can lead to situations in which the values of the variable to be synthesized are completely deterministic given the other variables. The CART synthesizer can get stuck in such a situation, simply replicating the records from the original data. While such a problem can easily be avoided by not using the fully conditional specification approach (the approach offers no advantages in the context of synthetic data), this example still highlights that it would be na\"ive to assume that fully synthetic data will never pose any threats of disclosing sensitive information. However, measuring these risks is challenging and research in this area is still limited.

We start this section by reviewing the approaches that have been proposed in the literature to assess risks for fully synthetic data. In principle, these measures can also be used to assess the risks for partially synthetic data, while the risk measures that we review in the second part of this section are only useful for partial synthesis as they try to assess the risk of re-identification for the generated data. We also refer the interested reader to \citet{hubayesian}, which contains a detailed review of Bayesian risk measures for synthetic data.

\subsection{Measuring the Risk of Disclosure for Fully Synthetic Data}
Even though the link between the original and the synthetic data is broken with full synthesis, some agencies still evaluate how many synthetic records have a unique match in the original data. The reasoning behind this evaluation is that the agencies are concerned about perceived risks. Survey respondents might be concerned if they find a synthetic record that exactly matches their own record, especially if their combination of attributes makes them unique in the original data.

Some authors \citep{park_data_2018,zhao_ctab-gan_2021} also compute the distance between the synthetic data records and their closest neighbors in the original data. The average of these distances across all synthetic records is then used as a risk measure. From a practical perspective, it is not obvious which risk this measure is supposed to quantify. Even if the average distance is small, the distance could be large for some records. A potential attacker would never know which records have small distance and even if the distance is small, this does not necessarily imply a risk if the closest record is in a high density area of the data distribution.

Another measure that evaluates risk by matching cases from the original and synthetic data was proposed by \cite{taub2019}. They suggest dividing the variables in the dataset into key variables, which are assumed to be known by the attacker, and target variables, which the attacker tries to infer. They assume that the attacker focuses on records with low $l$-diversity for the target variables within a given equivalence class given by the key variables. Let $K$ denote the vector containing the key variables and $T$ denote the vector of target variables. The authors define the Within Equivalence Class Attribution Probability (WEAP) as
\[
WEAP_j=Pr(T_j|K_j)=\frac{\sum_{i=1}^nI(T_i=T_j,K_i=K_j)}{\sum_{i=1}^nI(K_i=K_j)},
\]
where $I(\cdot)$ is the indicator function that is one  whenever the statement inside the parentheses is true and zero otherwise, and $n$ is the size of the database.
In their application, the authors focus on those synthetic records for which $WEAP_j=1$. For those records, they compute the Targeted Correct Attribution Probability (TCAP):
%\begin{eqnarray*}
\[
TCAP_{sj}%&=&
=Pr(T_{sj}|K_{sj})_o%\\
%&=&
=\frac{\sum_{i=1}^nI(T_{o,i}=T_{s,j},K_{o,i}=K_{s,j})}{\sum_{i=1}^nI(K_{o,i}=K_{s,j})},
\]
%\end{eqnarray*}

where the subscript $s$ denotes synthetic data and $o$ denotes the original data. The TCAP score is bounded between zero and one, with larger values indicating higher risks.

Another class of risk measures for fully synthetic data focuses on the fact that the synthesis models themselves can leak some information regarding the content of the original data. For example, when using a fully saturated log-linear model to synthesize a set of categorical variables combined with vague prior information, the existence of certain attribute combinations in the synthetic data reveals that the same combination must have been present in the original data. In the computer science literature, these types of risk evaluations are called membership attacks, as an attacker will learn that a certain record was present in the original data. Various strategies to estimate the risks from membership attacks have been proposed in the literature. Most of these approaches assume that the attacker already knows the true values for some target records and uses this information to learn whether these units are included in the original data \citep{stadler2021}. These evaluations are based on the strong assumption that the attacker is not interested in learning something new about a unit contained in the data. Instead, the only goal is to learn whether the unit was part of the original data. There are situations in which learning this information is considered unacceptable: some laws explicitly state that such risks must be avoided. In addition, sometimes the fact that someone is contained in a database already reveals sensitive information, if the database only contains a specific subgroup of the population such as the Survey of Prison Inmates conducted by the Bureau of Justice Statistics in the United States.

However, there are also risk measures based on inferential attacks that do not make such strong assumptions. Borrowing ideas from the differential privacy literature, \cite{reiter2014} propose strategies to compute the posterior distribution $f(Y_i|D,X, M, d_{org}^{-i})$, where $Y_i$ is the original value of some variable $Y$ for unit $i$, $D$ is the synthetic data, $X$ might contain unchanged values from the original data ($X$ will be empty for full synthesis), $M$ contains information about the synthesis model and $d_{org}^{-i}$ is the original data excluding record $i$. The approach evaluates, how much an attacker can learn about an unknown value $Y_i$ after seeing the synthetic data. If the posterior distribution for $Y_i$ has low variability (especially if compared to the prior distribution before seeing the synthetic data) disclosure can occur. In principle, the strong assumption that the attacker knows all the information from the original data except for record $i$ is not strictly necessary. However, in practice, it is typically unavoidable to make the problem computationally tractable. But even with these assumptions, this risk assessment is only feasible if the number of variables in the data is very limited (see \cite{HuReiterWang2014} and \cite{mcclure2016assessing} for illustrations).

In general, measuring disclosure risks for fully synthetic data remains challenging. While most researchers agree that fully synthetic data are not free from risk, more research is needed to quantify these risks under realistic settings.

\subsection{Measuring the Risk for Partially Synthetic Data}
As indicated above, most of the risk measures from the previous section can also be used for partial synthesis. However, the fact that synthetic records are only generated for units that were already included in the original data implies that each record in the synthetic data has a unique match in the original data. Thus, one way to measure the risk with partially synthetic data is to evaluate whether an attacker would be able to re-identify some records in the synthetic data. Building on previous work in \cite{reiter2005}, \cite{reiter:mitra:09} developed strategies to measure the risk of re-identification for partially synthetic data.

Borrowing from \cite{DrechReit10}, the risk computations can be summarized as follows. Suppose the intruder has a vector of information, $\mathbf{t}$, on a particular
target unit in the population $\mathbf{P}$.
Let $t_0$ be the unique identifier of the target,
and let $P_{i0}$ be the (not released) unique identifier
for record $i$ in $\mathbf{d}_{syn}$, where $\mathbf{d}_{syn}$ denotes the synthetic data and $i=1,\dots,n$.
Let $\mathcal{S}$ be any information released about the synthesis models.

The intruder's goal is to match unit $i$ in $\mathbf{d}_{syn}$
to the target when  $P_{i0}=t_0$. Let $J$ be a
random variable that equals $i$ when $P_{i0}= t_0$ for
$i \in \mathbf{d}_{syn}$. The intruder thus seeks to calculate
$Pr(J=i |\mathbf{t},  \mathbf{d}_{syn}, \mathcal{S})$
for $i=1,\dots, n$. Because the intruder does not know the
actual values of the synthesized variable $Y^*$, he or she should integrate over
its possible values when computing the match probabilities. Hence,
for each record he or she computes

%\begin{align} \label{eq:risk}
%\begin{split}
%&
\[\label{eq:risk}
Pr(J=i |\mathbf{t}, \mathbf{d}_{syn}, \mathcal{S}) =   %\\ & \int Pr(J=i | \mathbf{t},
\mathbf{d}_{syn}, Y^*, \mathcal{S}) Pr(Y^*|\mathbf{t}, \mathbf{d}_{syn}, \mathcal{S}) d Y^*.
\]
%\end{split}
%\end{align}

This construction suggests a Monte Carlo approach to estimating
each $Pr(J=i |\mathbf{t}, \mathbf{d}_{syn}, \mathcal{S})$. First, sample a
value of $Y^*$ from $Pr(Y^*|\mathbf{t},
\mathbf{d}_{syn}, \mathcal{S})$. Let $Y_{new}$ represent one set of
simulated values. Second, compute
$Pr(J=i|\mathbf{t}, \mathbf{d}_{syn}, Y^*=Y_{new}, \mathcal{S})$
using a matching strategy such as nearest neighbor matching assuming $Y_{new}$ are
collected values. This two-step process is iterated $h$
times, where ideally $h$ is large, and (\ref{eq:risk}) is estimated as
the average of the resultant $h$ values of $Pr(J=i|\mathbf{t},
\mathbf{d}_{syn}, Y^*=Y_{new}, \mathcal{S})$. When $\mathcal{S}$ has no
information,  the intruder treats the simulated values as
plausible draws of $Y^*$.

The disclosure risk can be measured using summaries of these identification probabilities. It is reasonable to assume that the intruder
selects as a match for $\mathbf{t}$ the record $i$ with the highest value
of $Pr(J=i|\mathbf{t}, \mathbf{d}_{syn}, \mathcal{S})$, if a unique maximum exists.
\cite{reiter:mitra:09} proposed three risk measures: the expected match risk, the true match rate, and the false match rate.
Let $c_i$ be the number of records
with the highest match probability for the target $\mathbf{t_i}$; let $I_i=1$ if the true match is among the $c_i$
units and $I_i=0$ otherwise. The expected match risk equals $\sum I_i/c_i$. When $I_i=1$ and $c_i>1$, the contribution of
unit $i$ to the expected match risk reflects the intruder randomly
guessing at the correct match from the $c_i$ candidates. Let $K_i=1$ when $c_iI_i=1$ and $K_i=0$
otherwise and let $N$ denote the total number of target records. The true match rate equals $\sum{K_i}/N$, which is the percentage of
true unique matches among the target records. Finally, let $F_j=1$
when $c_j(1 - I_j)=1$ and $F_j=0$ otherwise and let $s$ equal the
number of records with $c_i=1$. The false match rate equals $\sum
F_j/s$, which is the percentage of false matches among unique matches. Risk measures inspired by this methodology are available in the R package IdentificationRiskCalculation \citep{hornbyidentification}.

%ALTERNATIVELY, WE COULD USE THIS MUCH SHORTER SUMMARY OF THE RISK MEASURES:
%Borrowing from CITE:DRECHSLER:HU, suppose the intruder has information on some target records which she will use in a record linkage attack to identify the targets in the released data. Similar to the concept of probabilistic record linkage, the idea is to estimate the probability of a match between each target record and each record in the released file. The record with the highest average matching probability across the synthetic datasets is the declared match.

%The authors propose three risk measures that are summaries of these matching probabilities. The expected match risk, which computes the expected number of correctly declared matches, the true match rate, which computes the number of correct single matches among the target records, and the false match rate, which computes the number of erroneously declared single matches among all declared single matches.

These risk assessments are based on the conservative assumption that the intruder knows that the target record is included in the released data. Extensions of the approach which also account for the extra uncertainty from sampling if the intruder does not know whether the individual he or she is looking for participated in the survey are given in \cite{dre:rei:08}.

\section{Conclusion}
The interest in synthetic data has been growing steadily over the last thirty years. While the focus was on methodological aspects and statistical properties during the first decade, first applications started to appear around the turn of the century. The great success of GANs, which always require generating synthetic data even if the final goal is not to disseminate these data, had a huge impact on the synthetic data movement, especially in the computer science community. The availability of easy to use software such as synthpop \citep{raab2016} or the synthetic data vault \citep{patki2016synthetic} also meant that more statistical agencies and other data disseminating organizations were able to explore the approach without the need to implement the synthesizers from scratch.

In this paper, we reviewed the historic developments of the synthetic data approach, offered a taxonomy of approaches, and discussed methods to measure risk and utility of the generated data. For organizational reasons, we treated the statistical approach separately from the computer science approach. While it is true that the developments in the two fields mostly happened independently with little exchange between the disciplines, the lines have always been blurry (for example, \cite{abo:ger:08} already integrate ideas from both fields), and the increasing number of collaborations between statisticians and computer scientists in recent years will hopefully make this distinction obsolete in the future.

Furthermore, most of the applications of the synthetic data approach do not use the synthetic data as the final product. The synthetic data are either used for training purposes \citep{forbes2008} or to develop software code in preparation for working with the real data \citep{deWolf2015,raab2021}. Even in those cases in which final access to the real data is not possible, the data providers typically guarantee that they will run the final results on the original data and report back the results if they can be released without violating confidentiality \citep{benedetto2018, burman2019}. This implies that procedures for obtaining valid variance estimates from the synthetic data as discussed in Section \ref{sec:mi_inference} are less relevant in practice, and the fact that many of the computer science approaches never achieved this goal is less of a concern.

For those cases in which access to the original data cannot be provided, verification servers can be a useful alternative. These servers hold both the synthetic and the original data. Researchers can submit their analysis of interest to the server, it runs the analysis on both datasets, and reports back some fidelity measure how close the results from the synthetic data are to the results based on the original data. Compared to the guarantee of running the final models on the original data, verification servers have the advantage that the procedure can be automated. Since the server only reports a fidelity measure and not the actual results, no manual output checking is required. This means that the server could also be used frequently during data preparation and not only for the final model. However, some care must be taken, as even fidelity measures might spill sensitive information. Developing measures that are informative but at the same time are guaranteed not to spill sensitive information is an area of active research \citep{reiter2009,mcclure2012,barrientos2018,yu2018}.

A systematic comparison between the approaches developed in the different fields is currently lacking, although \cite{little2021generative} offers some first insights. The authors compared several synthesis strategies based on CART models, Bayesian Networks, and two GAN implementations (tableGAN and CTGAN). They found that the sequential-regression-based CART approach offered the highest utility, but also the highest risk. One of the GANs (tableGAN) resulted in unacceptably low utility, while CTGAN and the approach based on Bayesian Networks performed almost similarly. However, this evaluation was based on only one dataset and also relied on the default settings of the different software packages that were used for the different synthesis approaches. More extensive evaluations of the advantages and disadvantages of the various approaches that have been proposed in the literature would be an important area of future research.

\section*{Acknowledgements}
This work was partially supported by the German Federal Institute for Drugs and Medical Devices and a grant from the German Federal Ministry of Education and Research (grant number 16KISA096). The authors are grateful for helpful feedback on an earlier version of this paper from the FK2RG group at Mannheim Unversity and LMU Munich.

\bibliographystyle{chicago}
\bibliography{biblitreviewdatasyn.bib}      % Bibliography file (usually '*.bib')

%% or include bibliography directly:
% \begin{thebibliography}{}
% \bibitem{b1}
% \end{thebibliography}

\end{document}